\documentclass[aip,jcp,reprint,twocolumn,amsmath,amssymb,floatfix,citeautoscript]{revtex4-1}
\usepackage{graphicx}
\renewcommand{\vec}[1]{\mathbf{#1}}

\begin{document}

\title{A small subset of normal modes mimics the properties of dynamical heterogeneity in a model supercooled liquid}

\author{Glen M. Hocky}
\affiliation{Department of Chemistry, Columbia University, 3000 Broadway, New York, New York 10027, USA}
\author{David R. Reichman}
\email{drr2103@columbia.edu}
\affiliation{Department of Chemistry, Columbia University, 3000 Broadway, New York, New York 10027, USA}

\begin{abstract}
In this work, we study the nature of transitions between inherent structures of a two-dimensional model supercooled liquid. We demonstrate that these transitions occur predominately along a small number of directions on the energy landscape. Moreover, we show that the number of such directions decreases as the temperature of the liquid is decreased in the mildly supercooled regime, in concert with earlier studies on an athermal jamming system. We show that this decrease happens in parallel with a change in character of the transitions as dynamics in the system become more heterogeneous and localized. We investigate the origin of these trends, which suggests interesting connections between jamming and thermal glassy phenomena.
\end{abstract}

\maketitle

\section{Introduction}
Understanding the formation of glass from a slowly cooled liquid remains one of the great unsolved problems in condensed matter science\cite{Ediger-JPC1996,Debenedetti-Nature2001,Berthier-RMP2011}. The lack of an obvious change in symmetry and the fact that the transformation does not occur at a well-defined thermodynamic critical point renders the problem of vitrification more difficult to describe than crystallization. Despite this challenge, many aspects of glass formation appear to be generic and not tied to the specific details of the liquid under investigation. In particular, the phenomenology of dynamical heterogeneity\cite{Ediger-ARPC2000}, which includes growing dynamical length scales, violations of the Stokes-Einstein relation, and prominent non-Gaussian displacements in the tails of real-space distribution functions, provides a framework upon which theories of the glass transition must be based\cite{Berthier-Book2011}.

A problem that bares some similarity with the standard laboratory glass transition is the jamming transition of hard spheres\cite{Liu-Nature1998,Berthier-RMP2011,Berthier-Book2011,Liu-ARCMP2010}. Here, slow compression of the system may lead to the formation of a disordered solid.  This problem is simpler than the vitrification of typical liquids in the sense that a single control variable, the packing fraction, unambiguously tunes the transition upon approach to an amorphous configuration with a maximally allowed density.  While the relationship between the standard glass transition and the jamming transition is currently vigorously debated \cite{Zhang-Nature2009,Haxton-PRE2011,Ikeda-PRL2012}, it is clear that many aspects of the behavior of supercooled liquids and suspensions close to the jamming transition share important similarities. In particular, jamming systems display the hallmarks of dynamical heterogeneity first exposed in finite temperature fluids\cite{Dauchot-PRL2005,Keys-NatPhys2007,Dauchot-ChapterBook2011}.

One approach that has been useful in both the study of supercooled liquids and the jamming transition invokes the notion of soft modes \cite{Silbert-PRL2005,Xu-PRL2007,Xu-PRL2009,Goodrich-PRL2012}. Soft modes represent a subset of the low frequency harmonic or quasi-harmonic displacements of particles around metastable configurations \cite{Schober-PRB1991,Laird-PRL1991}. Within traditional mean-field theories of glassy liquids, soft modes characterize the motion that involves relaxation of groups of particles when marginally metastable states first appear \cite{Biroli-PRL2006,Krzakala-PRE2007,Kurchan-JPA1996,Berthier-RMP2011}. Jamming systems provide a concrete case where such modes play a prominent role. In particular, numerical simulations provide clear indications of {\em diverging length scale} connected to the jamming transition associated with soft modes\cite{Silbert-PRL2005}. In addition, soft modes appear to play a critical role in defining the mechanical stability of athermal packings close to jamming\cite{Brito-EPL2006,Brito-JCP2009}.  Currently the evidence for a primary role played by soft modes in the jamming transition is more extensive than it is for vitrification in thermal systems.

Despite the aforementioned lack of clarity of the role played by soft modes in supercooled liquids, interesting correlations between dynamics and soft modes in such systems have been uncovered via computer simulations. In a pioneering set of studies, Harrowell and coworkers defined the ``isoconfigurational ensemble'' as a means of quantitatively assessing the influence of structure on subsequent dynamics\cite{WidmerCooper-PRL2004,WidmerCooper-JPCM2005,WidmerCooper-PRL2006}. These studies revealed that regions that were quantifiably ``softer'' than average were more likely to be involved in large amplitude dynamically heterogeneous motion when averaged over many independent simulations \cite{WidmerCooper-PRL2004,WidmerCooper-JPCM2005,WidmerCooper-PRL2006}. Later it was demonstrated that the real space properties of the inherent structure normal modes correlate closely with dynamical heterogeneities and irreversible configurational rearrangements\cite{WidmerCooper-NatPhys2008,WidmerCooper-JCP2009,Candelier-PRL2010,Xu-EPL2010}.  Such connections have also been demonstrated in experiments performed on colloidal suspensions \cite{Brito-SoftMat2010,Ghosh-PRL2010,Chen-PRL2010,Chen-PRL2011}.  It remains unclear in the examples discussed above if the harmonic soft modes are an active player in the dynamics or if they are spectators whose positions correlate with regions of dynamical activity but which do not influence particle relaxation.

With an appropriate definition of normal modes emerging from the free energy landscape of hard spheres, Brito and Wyart have shown that dynamical heterogeneity in jamming systems also correlates with low frequency modes \cite{Brito-EPL2006}.  They have demonstrated that relaxation events in simulated hard spheres proceed along a small number of mode directions. Further, they have explicitly demonstrated that the number of such directions systematically decreases as the jamming threshold is approached \cite{Brito-JSM2007, Brito-JCP2009}.  The purpose of this work is to explore the possibility that a similar effect may occur in {\em thermal} systems as temperature is lowered, even in the absence of an analog of the jamming threshold where the correlation length associated with soft modes diverges.  

The paper is organized as follows. In Section \ref{sec:model} we will present the details of the two-dimensional glass forming liquid that we study here. In Section \ref{sec:InherentDynamics} we will review the notion of inherent structures (IS) and IS trajectories, and present relevant properties of the IS trajectories for our model. In Section \ref{sec:modes} we will discuss normal modes (NM), the NM properties of our system, and the decomposition of IS transitions onto a basis of modes. In Section \ref{sec:results} we will discuss the quantitative results of the decomposition procedure just described, and finally in Section \ref{sec:conclusion} we will summarize our results and frame them in a broader context.

\begin{figure}
\includegraphics[scale=0.72]{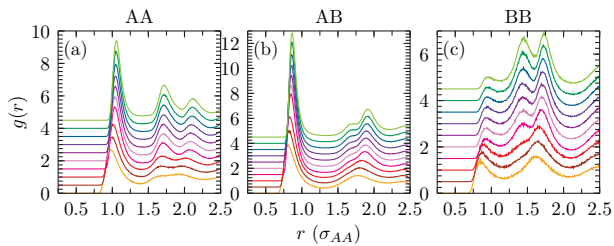}
\caption{Radial distribution functions for the 2D 65:35 Kob-Andersen Lennard-Jones system at $\rho=1.2$. Temperatures are $T=\{5.0,3.0,2.0,1.0,0.9,0.8,0.7,0.6,0.5,0.45\}$ from bottom to top. Each $g(r)$ is shifted up by 0.5 from that of the preceding temperature for clarity.}
\label{fig:rdf}
\end{figure}

\section{Model details}
\label{sec:model}
\begin{figure}
\includegraphics[scale=0.8]{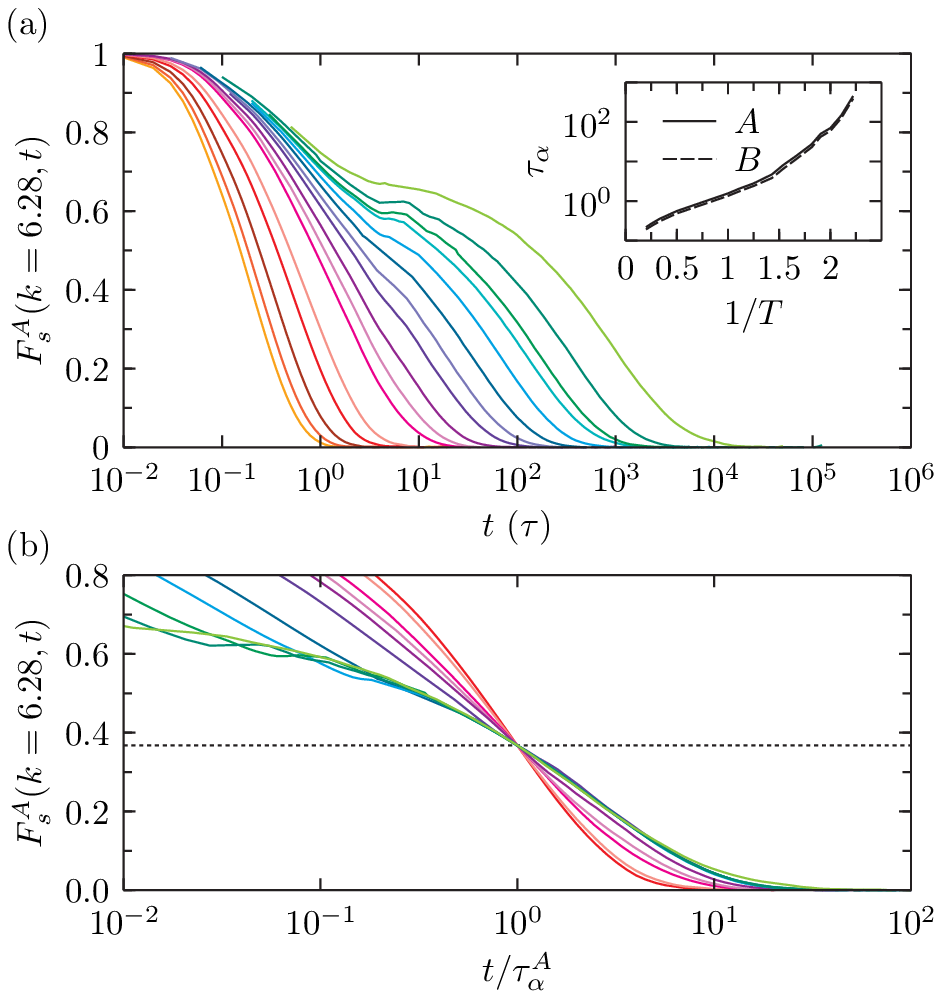}
\caption{ (a) Self-intermediate scattering function for $A$ type particles from $NVE$ molecular dynamics at the temperatures listed in Table ~\ref{tab:taualpha}, appearing from left to right. Inset: Alpha-relaxation times for both $A$ and $B$ type particles, with data given in Table ~\ref{tab:taualpha}. (b) $F_s^A(k=6.28,t)$ with the horizontal axis scaled by the alpha relaxation time $\tau_\alpha^A$. The curves collapse in the $\beta$-regime (the shortest times shown, where the correlation function has plateaued) as the temperature is lowered. The highest three temperatures are not shown. The dotted line shows the value $1/e$.}
\label{fig:dynamics}
\end{figure}

We study the two dimensional ``65:35'' Kob-Andersen binary Lennard-Jones mixture \cite{Bruning-JPCM2009}. In brief, it is characterized by the parameters $\sigma_{AB} = 0.8 \sigma_{AA}$, $\sigma_{BB} = 0.88 \sigma_{AA}$, $\epsilon_{AB} = 1.5 \epsilon_{AA}$, and $\epsilon_{BB} = 0.5 \epsilon_{AA} $, with $65\%$ of the particles being of type A. All particles have unit mass. Time scales are reported in Lennard-Jones units with $ \tau = \sqrt{ m \sigma_{AA}^2 / \epsilon_{AA} } $ and we fix the number density at $\rho = 1.2$. All interactions were shifted and cut off at $r_{ij}=2.5\sigma_{ij}$. 
This model has been used previously because it resists crystallization in two dimensions better than the standard 80:20 variant\cite{Bruning-JPCM2009,Karmakar-PRE2012}. 
To our knowledge, structural and dynamical quantities for this system have not been previously reported. Here, for completeness, we briefly discuss these properties for $N=1000$. In Fig.~\ref{fig:rdf} we present the radial distribution functions for a series of temperatures that span the high and supercooled temperature regimes. In Fig.~\ref{fig:dynamics}(a) we show the self-part of the intermediate scattering function for the A-type particles, $F^A_s(k=6.28,t)$, as well as the $\alpha$ relaxation times as defined by $F_s(k=6.28,t=\tau_{\alpha})=1/e$. These values are also reported in Table ~\ref{tab:taualpha}. Finally, in Fig.~\ref{fig:dynamics}(b) we also show that time-temperature superposition appears to hold in the $\beta$-regime in a manner comparable to the 3D, 80:20 variant of Ref.~\onlinecite{Kob-PRL1994}. 

\begin{table}
\begin{tabular}{  c  c  c  c  c @{\hskip 0.2in} c  c  c  c  c  }
\hline
\hline
$T$ & $\tau_\alpha^A$ & $\tau_\alpha^B$ & $\tau_\alpha$ & $dt$ & $T$ &  $\tau_\alpha^A$ & $\tau_\alpha^B$ & $\tau_\alpha$ & $dt$\\
\hline
5.000 & $ 0.22 $ & $ 0.19 $ & $ 0.20 $ & 0.001 & 0.700 & $ 4.70 $ & $ 3.81 $ & $ 4.21 $ & 0.003 \\
4.000 & $ 0.27 $ & $ 0.23 $ & $ 0.25 $ & 0.001 & 0.650 & $ 8.10 $ & $ 6.51 $ & $ 7.35 $ & 0.003 \\
3.000 & $ 0.36 $ & $ 0.31 $ & $ 0.33 $ & 0.001 & 0.600 & $ 13.9 $ & $ 11.5 $ & $ 12.7 $ & 0.003 \\
2.000 & $ 0.56 $ & $ 0.49 $ & $ 0.52 $ & 0.001 & 0.550 & $ 27.5 $ & $ 22.9 $ & $ 25.3 $ & 0.004 \\
1.500 & $ 0.79 $ & $ 0.70 $ & $ 0.73 $ & 0.002 & 0.525 & $ 50.1 $ & $ 42.1 $ & $ 46.0 $ & 0.004 \\
1.000 & $ 1.56 $ & $ 1.35 $ & $ 1.45 $ & 0.002 & 0.500 & $ 70.1 $ & $ 59.3 $ & $ 64.8 $ & 0.005 \\
0.900 & $ 2.06 $ & $ 1.76 $ & $ 1.91 $ & 0.002 & 0.475 & $ 145  $ & $ 124  $ & $ 135  $ & 0.005 \\
0.800 & $ 2.80 $ & $ 2.40 $ & $ 2.60 $ & 0.002 & 0.450 & $ 449  $ & $ 388  $ & $ 418  $ & 0.005 \\ 
\hline                                 
\hline                                 
\end{tabular}
\caption{Alpha relaxation times ($\tau_{\alpha}$) for the 2D, 65:35 Kob-Andersen system, with $N=1000$. $\tau_{\alpha}$ is reported for $A$ and $B$ type particles, as well as for all particles together. Also listed for reference are integration time-steps used both in annealing the configurations and in generating inherent structure trajectories.\label{tab:taualpha}
}
\end{table}

\begin{figure*}
\includegraphics[scale=1.1]{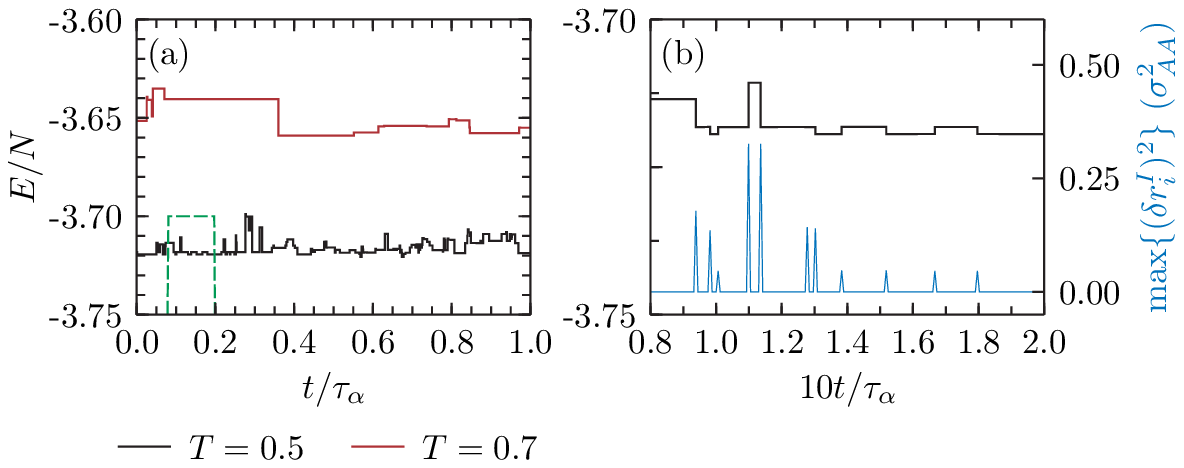}
\caption{(a) Inherent structure energies (per particle) at $T=0.7$ (above) and $T=0.5$ (below), plotted as a function of time scaled by the alpha relaxation time $\tau_\alpha$ at that temperature. Transitions occur between extended plateaus at both temperatures; they occur less frequently on an absolute time scale at the lower temperature but more frequently relative to the alpha relaxation time. The dashed box shows the area plotted in (b). (b) Above is a magnified look at the transitions at $T=0.5$, with energies per particle shown on the left. Below are the maximum squared displacements between subsequent inherent structures, with values shown on the right. Note every energy transition has a structural difference as measured by maximum squared displacement, and there is an additional transition at $10t/\tau_\alpha\approx1.28$ not visible in the energy trace. }
\label{fig:transitions}
\end{figure*}
\section{Inherent dynamics}
\label{sec:InherentDynamics}

The motion of particles in a supercooled liquid is extremely complex, and the ability of a single particle to change its current position is dependent on the interplay of fluctuations on many length scales. The notion of inherent structures (IS) has been developed as an aid to simplify the description of both structural and dynamical processes in supercooled liquids \cite{Stillinger-PRA1982,Weber-JCP1984,Weber-PRB1985}.
In the IS framework, the motion of the system as a whole is decomposed into transitions between local minima on the global potential energy surface (the ``inherent structures'') and the vibrations around these configurations. Each glassy configuration is then associated with the configuration which is the ``closest'' minimum, analogous to how one might associate a solid's configuration with a periodic crystal structure, ignoring thermal fluctuations. Here, ``closest'' is a practical definition meaning the structure obtained by a local energy minimization technique. A temporal series of IS generated by quenching a trajectory from a molecular dynamics simulation is termed an IS trajectory. Studies of these trajectories have provided interesting insights into the {\em true} dynamics in glassy systems, for example, revealing the presence and nature of string-like cooperative motion \cite{Sastry-Nature1998,Schroder-JCP2000,Keys-PRX2011}.

For this study, we will require IS trajectories at a series of decreasing temperatures. Specifically, configurations of 250 particles each were generated using $NVT$ dynamics in LAMMPS by annealing through the series of temperatures in Table ~\ref{tab:taualpha}, simulating for over $100 \tau_{\alpha}$ (as calculated for $N=250$) at the lower temperature for each subsequent annealing step\cite{Plimpton-JCP1995}.
Each resulting configuration was then simulated for an additional $100 \tau_{\alpha}$, and $F_s(k,t)$ was calculated for that entire trajectory, as well as for the first and second halves. At the lower temperatures, signs of aging were obvious in the majority of these simulations; only those configurations resulting in a simulation that showed little or no difference in dynamics between the first and second half of the trajectory were used for subsequent analysis. At all temperatures, we obtained at least 20 independent configurations for future study. The integration time steps used throughout this work are reported in Table~\ref{tab:taualpha}.

For each of these independent configurations, IS trajectories were generated by quenching configurations from molecular dynamics simulations using steepest descent minimization. Our system is quite small, rendering the IS meaningful in concept and easy to identify. Specifically, trajectories of length 100,000 integration steps were generated at each temperature, saving every configuration, except at $T=0.45$ where it was necessary to run 200,000 and save every 2 steps to obtain a comparable level of statistical accuracy. We generated IS trajectories using a bisection technique similar to that of Ref.~\onlinecite{Schroder-JCP2000}. A small number of configurations evenly spread throughout the trajectory were quenched to a local minimum. If two adjacent configurations differed (see next paragraph for more details), then we bisected that time interval and quenched the configuration halfway between the two endpoints. This procedure was iterated until a sequence of configurations was generated in which all transitions between minima occurred at the time scale of a single integration time step.

For every configuration in the new IS trajectory, we monitored both the energy and the displacement from the previous configuration. We calculated both the mean square displacement $ \langle \Delta \vec{R}_{IS}^2(t_i) \rangle = \frac{1}{N} \sum_{j=1}^N | \vec{r}^I_j(t_i)-\vec{r}^I_j(t_{i-1}) |^2 $ and maximum squared displacement $\max\{(\delta \vec{r}^I_i)^2\}  = \max_{1\le j\le N} \{| \vec{r}^I_j(t_i)-\vec{r}^I_j(t_{i-1}) |^2 \}$ where $t_i$ is the time index of the $i$'th inherent structure, $\vec{r}^I_j(t_i)$ is the position of particle $j$ in inherent structure $i$,  and $N$ is the total number of particles. We find that these two quantities are correlated and significant changes in one always occur in the other simultaneously.

A plot of the IS energy, as displayed in Fig.~\ref{fig:transitions}(a), reveals a series of plateaus as in previous works \cite{Schroder-JCP2000,Doliwa-PRE2003a,Doliwa-PRE2003b,Denny-PRL2003}. Differences in energy within these plateaus are on the order of $10^{-6}$ per particle or smaller. However, as discussed by Schr{\o}der, {\em et al.}, there is the small probability that a change in structure occurs between two configurations with a vanishingly small difference in energy\cite{Schroder-JCP2000}. Hence, as in that previous work, we will identify changes in inherent structure by the difference in positions and not by energy. We used  $\max\{(\delta \vec{r}^I_i)^2\}$ for this purpose but emphasize that choosing the mean square displacement would produce identical results. For the model studied here, we find that maximum squared displacements in the range of $10^{-5}$ to $10^{-2}$ are very rare. We therefore empirically choose a threshold value within this range and deem that whenever $\max\{(\delta \vec{r}^I_i)^2\} > 0.001$, a transition has occurred. The correlation between changes in structure and changes in energy is illustrated in Fig.~\ref{fig:transitions}(b).

It will be most important in future discussion to quantify the degree of localization of motion in an IS transition (or a normal mode, see Sec.~\ref{sec:modes}). The metric we shall use is the standard participation ratio. For a vector quantity $\vec{v}$ of length $d N$ such as the displacement field, the vector can be organized as a matrix $\mathbb{V}$ of size $N \times d$ where each row $1 \le i \le N$ contains a quantity pertaining to particle $i$. With these quantities, the participation ratio is defined as\cite{Laird-PRL1991},
\begin{equation}
P(\vec{v}) \equiv \left [ N \sum_{i=1}^N (\mathbb{V}_i \cdot \mathbb{V}_i )^2 \right ]^{-1}.
\label{eq:participation}
\end{equation}

\begin{figure}
\includegraphics[scale=0.8]{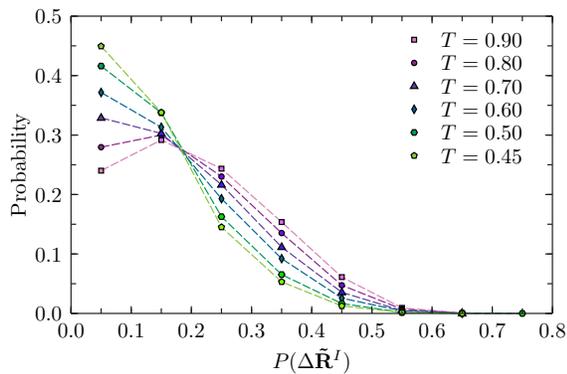}
\caption{At each temperature studied, the proportion of inherent structure transitions histogrammed into bins of size $P(\Delta \tilde{\vec{R}}^I )=0.1$. At high temperatures there are significantly more ``extended'' transitions than at low temperatures. Dashed lines are a guide to the eye. Note that calculations  on transitions with $P(\Delta \tilde{\vec{R}}^I ) \geq 0.55$ such as those in Figs.~\ref{fig:fhalf_disppart} and \ref{fig:contrib_modes_freqs} may be skewed due to poor statistics, but this does not affect any of our conclusions.}
\label{fig:disppart_count}
\end{figure}

We use this definition to calculate the participation ratio of transitions $P(\Delta \vec{\tilde{R}}^I (t_i))$ at each temperature. Nearly all participation ratios fall in a range from $0$ to $0.6$, with just a few occurring with values larger than $0.6$. As a matter of nomenclature only, we will refer to transitions with $P(\Delta \vec{\tilde{R}}^I (t_i))>0.3$ as ``extended'', and all others as ``localized''. A histogram is shown in Fig.~\ref{fig:disppart_count}. As the temperature is lowered, transitions with low participation ratio become much more prevalent i.e. motion is increasingly localized. Interestingly, the curves intersect for participation ratio $\approx 0.17$, although the meaning of this coincidence is unclear.
%perhaps due to an increasing barrier to collective rearrangements on lowering temperature expected for fragile glass formers by most theories of the glass transition \cite{Berthier-RMP2011}.
An illustration of the types of IS transitions is shown in Fig.~\ref{fig:displacements}.

\section{Normal modes}
\label{sec:modes}

Suppose a transition in a given IS trajectory is found to occur at time $t_i$. We wish to study the relationship between the normal modes of configuration  $\vec{R}_{IS}(t_i)$ and the transition $\vec{R}_{IS}(t_i) \rightarrow \vec{R}_{IS}(t_{i+1})$. To do this, we construct the Hessian or dynamical matrix for configuration $\vec{R}_{IS}(t_i)$ given by the matrix of second derivatives of the potential $V$,\cite{WidmerCooper-NatPhys2008}
\begin{equation}
\mathbb{H}(t_i)_{jk} = \frac{\partial^2 V( \vec{R}_{IS}(t_i) ) } { \partial \vec{r}^I_j \partial \vec{r}^I_k }.
\label{eq:hessian}
\end{equation}

\begin{figure}
\includegraphics[scale=0.9]{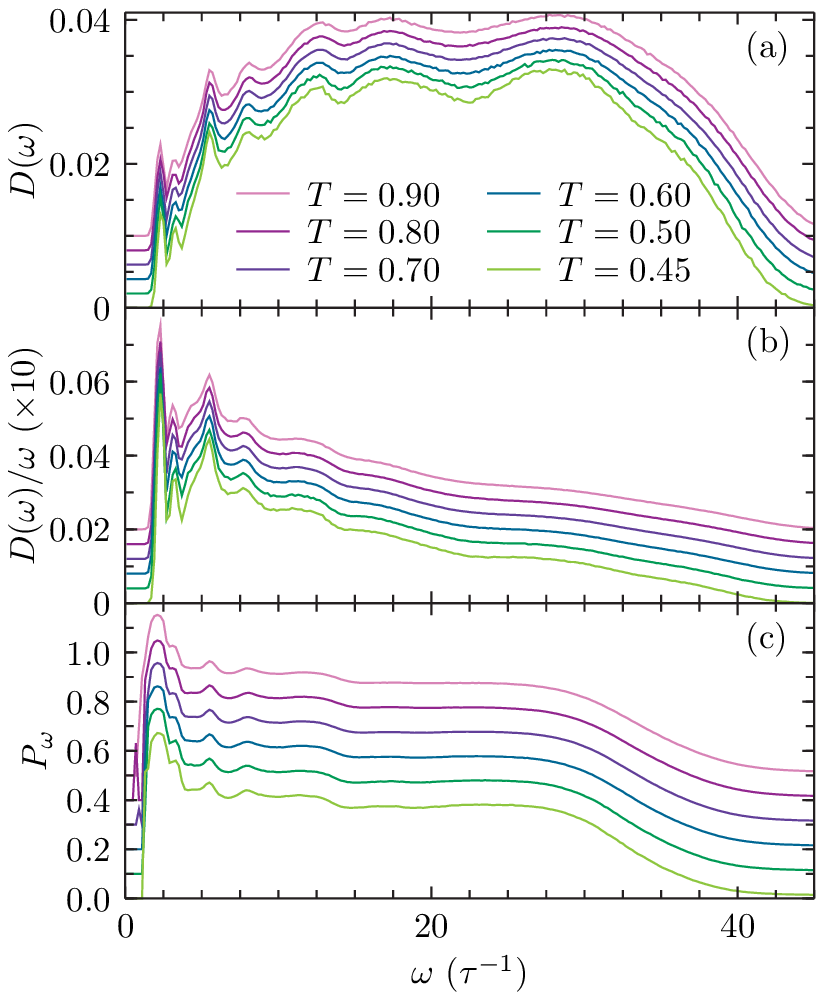}
\caption{(a) Density of states for the 2D Kob-Andersen model with $N=250$ at the series of temperatures shown. Curves are shifted up by 0.002 from the temperature below for clarity, the lowest temperature being at the bottom. (b) The same as (a) but the density of states has been divided by frequency, and scaled up by a factor of 10 for clarity. (c) Participation ratio as a function of frequency for the same series of temperatures as in (a). Curves are shifted up by 0.1 from the temperature below for clarity. Quantities in (a) and (b) are binned to a resolution of $0.2 \tau^{-1}$.}
\label{fig:dos}
\end{figure}

Normal modes are the eigenvectors obtained by diagonalization of this matrix. It is well known that the normal modes form a complete orthonormal basis for the motion of this configuration. For configuration $i$ the eigenvectors can be ordered by eigenvalue and will be denoted $\{\vec{e}_j^i\}$. For a $d$-dimensional configuration of $N$ particles, $\vec{e}_j^i$ will be a vector of length $d N$ and there will be $d N$ eigenvectors, of which $d$ will correspond to translational modes with eigenvalue $0$. The Hessian is real and symmetric and hence the eigenvalues and eigenvectors are real. 
The density of states (DOS) $D(\omega)$ is a histogram of frequencies obtained from the eigenvalue spectrum, and is plotted for our system at a series of temperatures in Fig.~\ref{fig:dos}(a). The same histogram is shown with the DOS divided by a factor of $\omega$ in Fig.~\ref{fig:dos}(b) so that one can identify the so-called ``boson peak'', where the number of low frequency modes exceeds the Debye prediction of $\omega^{d-1}$ \cite{Xu-PRL2007}.

The displacement vector for a transition is given by $| \Delta \vec{R}^I (t_i) \rangle = | \vec{R}_{IS}(t_{i}) \rangle - | \vec{R}_{IS}(t_{i-1}) \rangle $, and characterizes the inherent structure transition. For convenience, we will consider instead the normalized displacement vector defined as $| \Delta \vec{\tilde{R}}^{I} (t_i) \rangle = \Delta \vec{R}^{I} (t_i) / \sqrt{ \langle \Delta \vec{R}^{I} (t_i) | \Delta \vec{R}^{I} (t_i) \rangle } $. This can be projected onto a normal mode basis as,
\begin{equation}
| \Delta \vec{\tilde{R}}^I (t_i) \rangle = \sum_{j=1}^{d N} c^k_j(i) | \vec{e}_j^k \rangle .
\label{eq:decomposition}
\end{equation}

Note that we label the eigenvectors with configuration index $k$ to emphasize that these need not be the eigenvectors corresponding to time $t_i$. The coefficients $c^k_j(i)$ above are obtained through the orthonormality condition for the basis
\begin{equation}
\langle \vec{e}^k_j | \Delta \vec{\tilde{R}}^I (t_i) \rangle = \sum_{l=1}^{d N} c^k_l(i) \langle \vec{e}_j^k | \vec{e}_l^k \rangle = c^k_j(i).
\label{eq:coefficients}
\end{equation}

Since $ | \Delta \vec{\tilde{R}}^I (t_i) \rangle $ is normalized, its squared norm can be written as
\begin{equation}
1 = \langle \Delta \vec{\tilde{R}}^I (t_i) | \Delta \vec{\tilde{R}}^I (t_i) \rangle = \sum_{j=1}^{dN} |c^k_j(i)|^2.
\label{eq:completeness}
\end{equation}

We will refer to $|c^k_j(i)|^2$ as the {\em contribution} of mode $j$ of basis $k$ to transition $i$. These contributions can be ordered from highest to lowest in numerical value. 

In order to get an idea of how much each mode contributes to a transition, we consider the quantity $N_{1/2}$, which is defined to be the smallest number of modes necessary to reproduce $50\%$ of a given transition. We will also report $F_{1/2}$, the fraction of all modes necessary to reproduce $50\%$ of the motion, which has been used previously \cite{Brito-JSM2007,Brito-JCP2009}. In general we can also define the quantities $N_f$ and $F_f$ where $f$ is a fraction other than 0.5. To provide a unique definition of these quantities, we take $\{o_l^k(i)\}$ as a monotonically decreasing sequence of ordered contributions to transition $i$ from basis $k$ and then generate the cumulative series $s_m^k(i) =\sum_{l=1}^m o_l^k(i)$. Then
\begin{equation}
N^k_{f}(i) = \min\{ m | s_m^k(i) > f \}
\label{eq:fraction}
\end{equation}
and $F_f^k(i) = N_f^k(i)/(N d)$.

The definition of participation ratio given in Eq.~\ref{eq:participation} also applies to the normal modes obtained via diagonalization of the Hessian. By convention, the participation ratio averaged over modes with a given frequency is referred to as $P_\omega$. This quantity, which will play a role in our future discussion, is shown in  Fig.~\ref{fig:dos}(c) for a wide range of temperatures.

\begin{figure}
\includegraphics[scale=0.35]{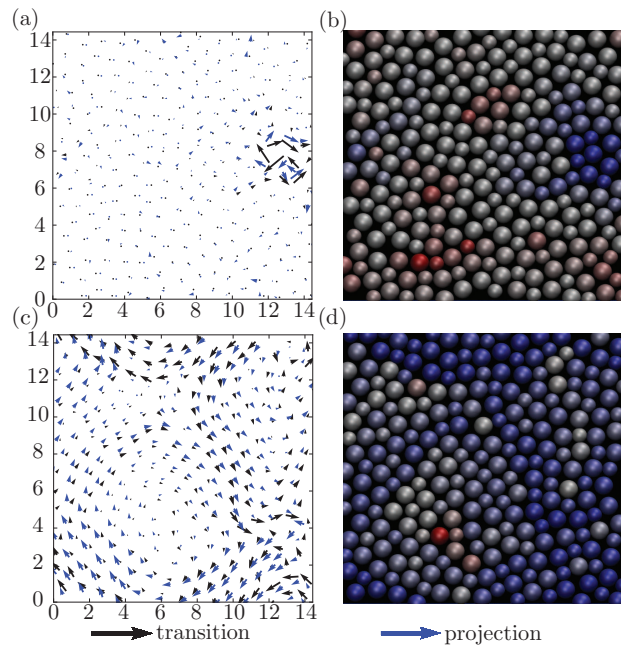}
\caption{Inherent structure transitions for the 2D Kob-Andersen model with $N=250$ at $T=0.5$. Axis labels are simulation coordinates. (a) A ``localized'' transition with participation ratio $P(\Delta \mathbf{\tilde{R}}^I)=0.04$ and $\langle (\Delta \vec{R}^I)^2\rangle=0.06$. Particle displacements are shown with black arrows and a projection of the top 37 highest contributing modes, reproducing $50\%$ of the transition is shown with blue arrows. Vectors are multiplied by a factor of two for clarity. (b) The same transition with the natural logarithm of square particle displacements $\ln((\delta \vec{r}^I)^2)$ represented from red at $-12.5$ to blue at $-0.7$. (c) Same as (a) but for an ``extended'' transition with  $P(\Delta \mathbf{\tilde{R}}^I)=0.5$ and $\langle (\Delta \vec{R}^I)^2\rangle=0.2$. Here only one mode is needed to project onto $60\%$ of the transition. (d) The same transition as (c), with $\ln((\delta \vec{r}^I)^2)$ ranging from $-9.43$ in red to $-1.48$ in blue.}
\label{fig:displacements}
\end{figure}

In Fig.~\ref{fig:displacements} we illustrate the concepts presented in this and the preceding section. Fig.~\ref{fig:displacements}(a) and (b) show an IS transition with low participation ratio, in which most of the contribution to the motion is localized to a few particles. In Fig.~\ref{fig:displacements}(c) and (d) we show an IS transition with high participation ratio, in which the motion is much more evenly spread throughout the system. In (a) and (c) we show the result of projecting the motion of an IS transition onto normal modes of the initial IS. The black arrows show the motion of the particles in the system during an IS transition, and the blue arrows show the sum of the normal modes that contribute to $N_{1/2}$ weighted by their coefficients as in Eq.~\ref{eq:decomposition}. 
We see that for both localized and extended transitions, the degree of similarity is striking. In the single example of an ``extended'' transition, a single normal mode projects onto $60\%$ of the actual displacement vector. We illustrate the opposite case with a particularly localized event, where just a few particles rearrange, and where a relatively large number of modes are needed to reproduce half of the motion compared to the average as discussed in Sec.\ref{sec:results}.
In what follows, we use the metrics $N_{1/2}$ and $F_{1/2}$ as a quantitative measure of how faithfully a given transition may be represented by a small set of normal modes.

\section{Results}
\label{sec:results}
\begin{figure}
\includegraphics[scale=0.8]{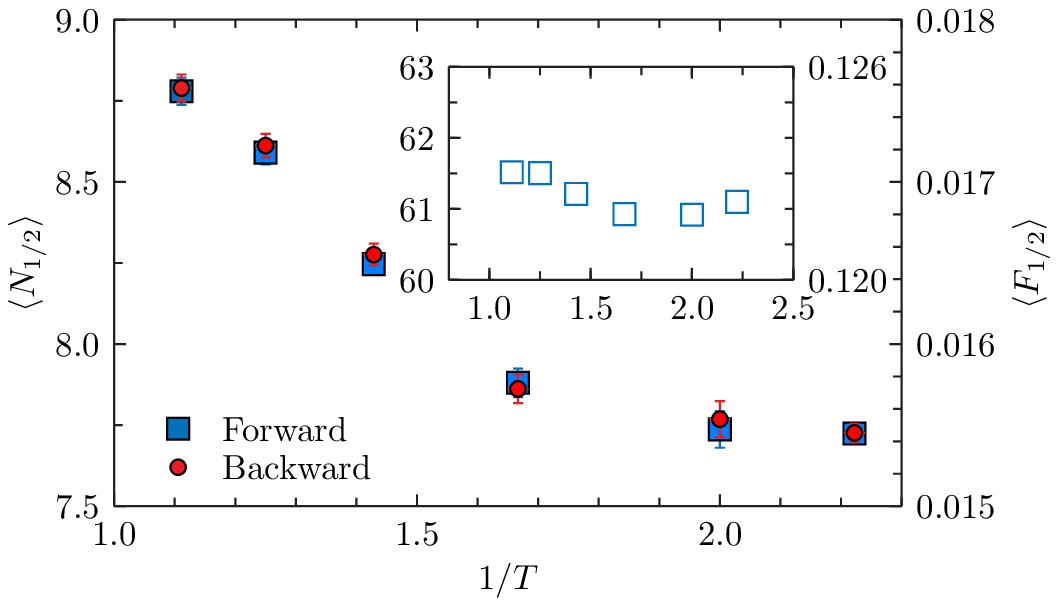}
\caption{ Average number of modes (left axis) or fraction of total modes (right axis) necessary to reproduce at least $50\%$ of total particle motion in inherent structure transitions at each temperature studied. Blue squares are for transitions in the ``forward'' direction and red circles in the ``backwards'' direction as described in the main text. Error bars shown are jackknife standard errors\cite{Efron-1993}. Inset: Symbols are the values of $\langle N_{1/2} \rangle$ and $\langle F_{1/2} \rangle$ where 2000 inherent structures at each temperature are projected onto the modes from each of twenty-five random configurations from an independent IS trajectory. Note the difference in magnitude between the vertical axis of the inset and the main figure.}
\label{fig:fhalftemp}
\end{figure}

\begin{figure}
\includegraphics[scale=0.8]{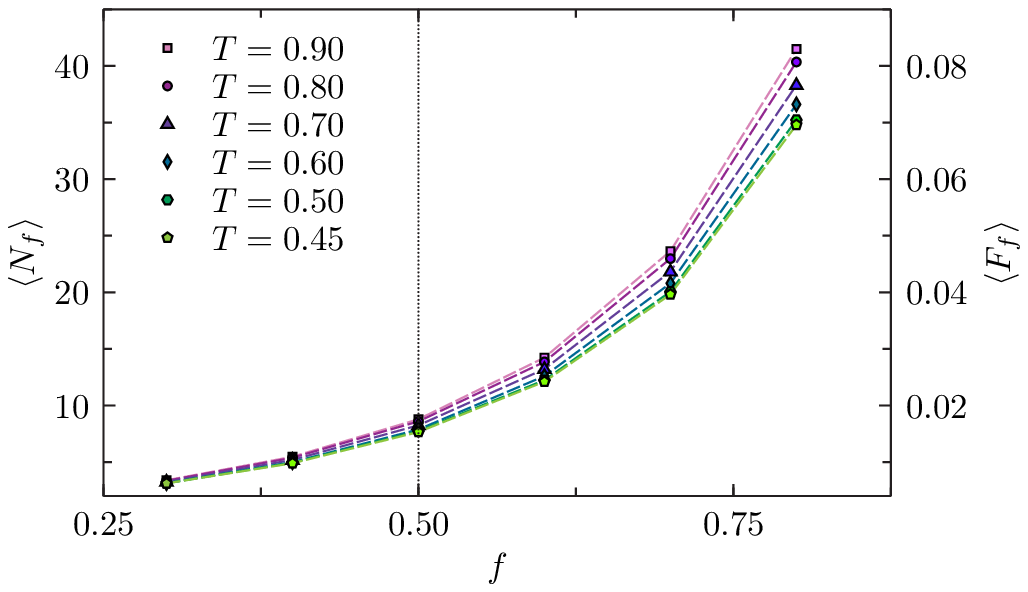}
\caption{Average number and fraction of total modes necessary to reproduce a fraction $f$ of inherent structure transitions from $f=0.3$ to $f=0.8$. A dotted line shows $f=1/2$, the value focused on for most of this work. A slice through this line gives the data shown in Fig.~\ref{fig:fhalftemp}. $\langle N_f \rangle $ decreases with temperature for all values of $f$. Jackknife error bars are approximately the size of the symbols\cite{Efron-1993}. Dashed lines are a guide to the eye.}
\label{fig:fall}
\end{figure}

\begin{figure}
\includegraphics[scale=0.8]{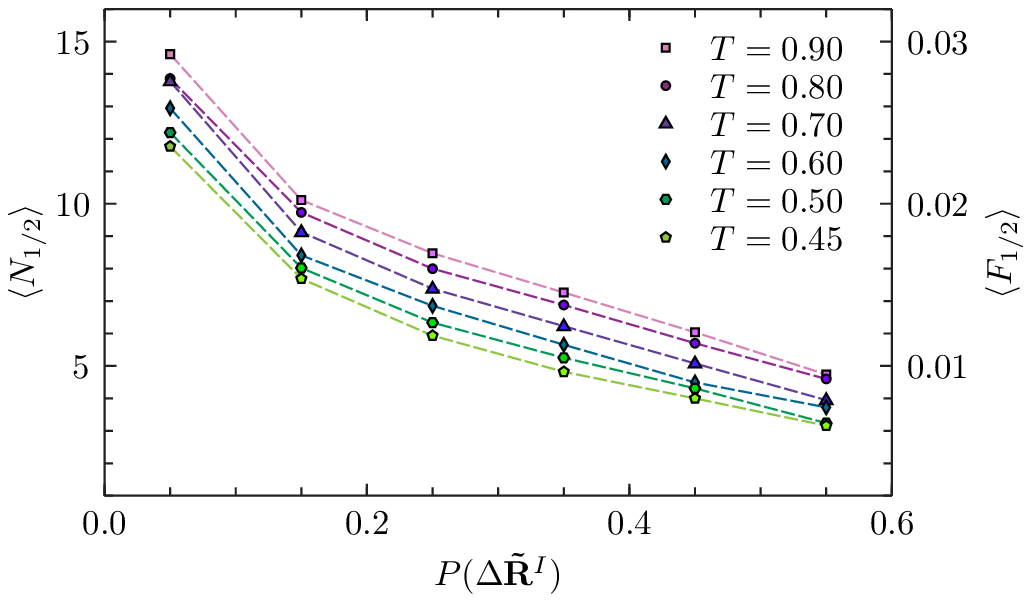}
\caption{ $\langle N_{1/2} \rangle $ and $\langle F_{1/2} \rangle $ for inherent structures as a function of participation ratio of IS transitions. Bins of size $P(\Delta \tilde{\vec{R}}^I )=0.1$ are used. ``Extended'' transitions (measured by $P(\Delta \tilde{\vec{R}}^I )$ require fewer modes to reproduce their transitions. $\langle N_{1/2} \rangle $ and $\langle F_{1/2} \rangle $ decrease uniformly for all values of $P(\Delta \tilde{\vec{R}}^I )$ as temperature is lowered. Jackknife error bars are again approximately the size of the symbols and dashed lines a guide to the eye\cite{Efron-1993}. Data for $P(\Delta \tilde{\vec{R}}^I )>0.55$ are not shown due to poor statistics, as seen in Fig.~\ref{fig:disppart_count}. }
\label{fig:fhalf_disppart}
\end{figure}

\begin{figure}
\includegraphics[scale=0.8]{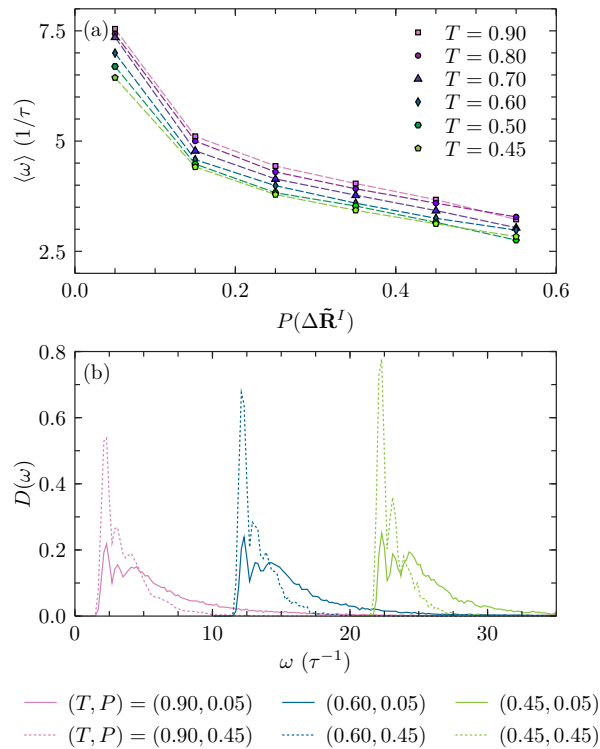}
\caption{(a) Average frequency of modes contributing to $\langle N_{1/2} \rangle $ as a function of IS transition participation ratio histogrammed into bins of size $P(\Delta \mathbf{\tilde{R}}^I)=0.1$. ``Extended'' transitions are made up of lower frequency modes. The frequency of contributing modes decreases with temperature for all sized transitions. Jackknife errors are smaller than the points shown \cite{Efron-1993}. Dashed lines are a guide to the eye. (b) Frequency distribution of modes used to compute $\langle \omega \rangle$ for three temperatures $T=\{0.9,0.6,0.45\}$, and two values of $P(\Delta \mathbf{\tilde{R}}^I)=\{0.05,0.45\}$ (referred to as $P$ here for convenience). Other temperatures and $P$ values interpolate between these curves. Temperatures are shifted to the right from the temperature above by $\omega=10$ and curves for $P=0.45$ are dotted. We see that curves are more sharply peaked for higher $P$ and the peak height increases as temperature is lowered. }
\label{fig:contrib_modes_freqs}
\end{figure}

\begin{figure}
\includegraphics[scale=0.75]{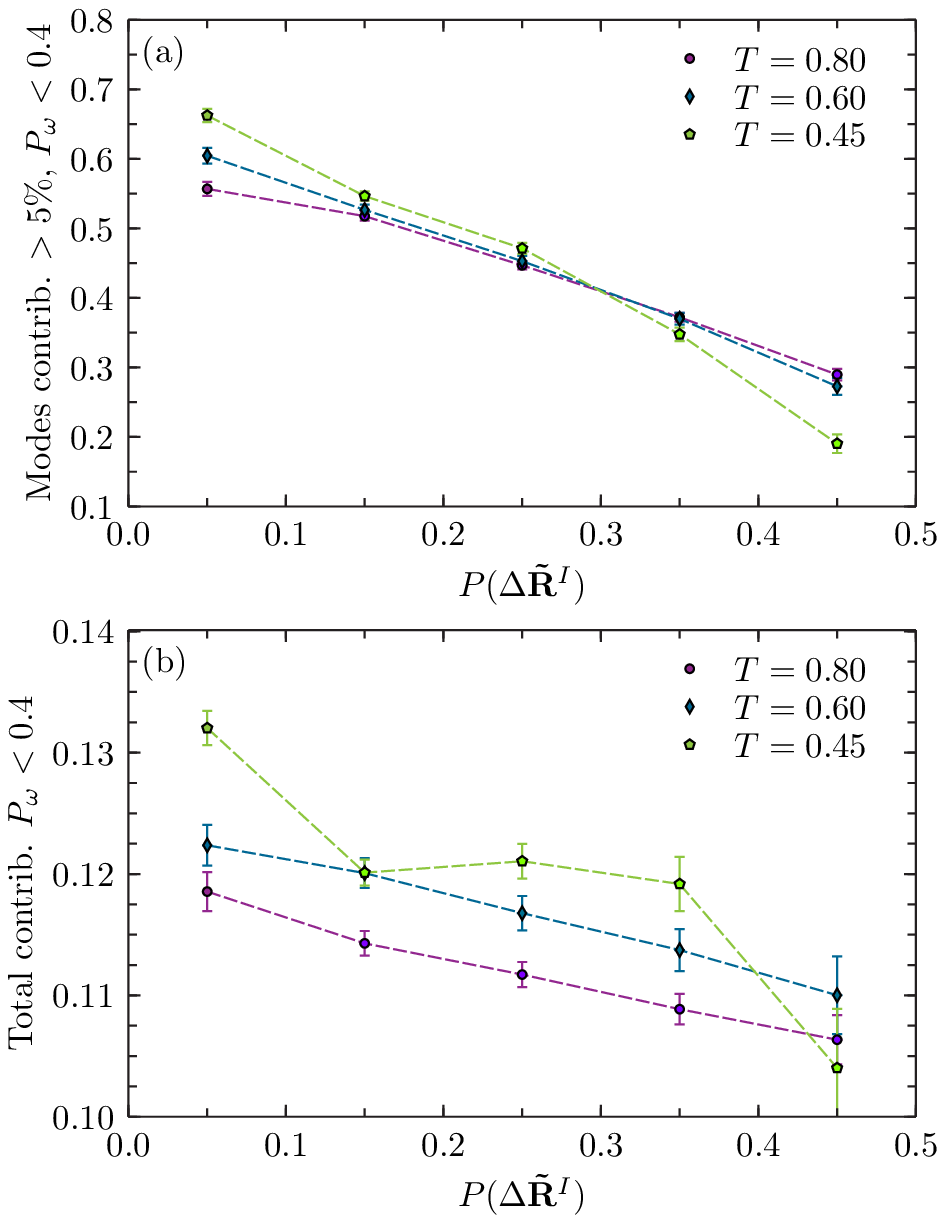}
\caption{(a) Average number of modes which have a participation ratio $P_\omega < 0.4$ and which contribute more than $5\%$ to transitions of a given type at three different temperatures. (b) For the same temperatures, average total contribution to a single transition of a given type from modes with $P_\omega < 0.4$. Error bars are jackknife standard errors \cite{Efron-1993}. Modes with low participation ratios contribute more to ``localized'' transitions ($P(\Delta \tilde{\vec{R}}^I)<0.3$) at low temperatures than at high temperatures. }
\label{fig:soft_contrib}
\end{figure}

We begin this section with a discussion of our main result. We examine the temperature dependence of the quantity $\langle N_{1/2} \rangle$ as defined in Section \ref{sec:modes} (where brackets denote average over many transitions) following Brito and Wyart, who showed that this quantity decreased systematically in their system of hard spheres as the jamming transition was approached \cite{Brito-JSM2007}. Analogously, we plot in Fig.~\ref{fig:fhalftemp} the identical quantity for our thermal system as a function of inverse temperature. It is clear that as our system becomes increasingly supercooled the number of normal modes needed to approximate the correlated atomic motion occurring in an inherent structure transition decreases. The degree to which this is true is independent of whether the pre- or post-transition inherent structure is used to produce the normal mode basis (which we term ``forward'' or ``backward'' transitions, respectively). This fact, which is required if we are to interpret the mode basis as a set of directions along which reversible motion occurs on a potential energy landscape, is not entirely trivial as the modes of the two configurations are distinct.\footnote{In fact, all quantities computed which compare modes with inherent structure transition displacement vectors were identical within error bars if computed using pre- and post-transition modes, and hence when calculating the values in Figs.~\ref{fig:fall}-\ref{fig:soft_contrib} we average both sets of data for increased statistical accuracy.}

To gain some insight into the magnitude of  $\langle N_{1/2}\rangle$, as well as the significance of the trend seen in Fig.~\ref{fig:fhalftemp}, we also reconstruct the dynamical activity exhibited during inherent structure transitions with a random set of modes obtained from distant inherent structures sampled at the same temperature as the transitions under consideration.  The inset of Fig.~\ref{fig:fhalftemp} shows that the value of  $\langle N_{1/2}\rangle$ obtained from this procedure exceeds that obtained with modes associated with the actual transition by approximately one order of magnitude at the temperatures considered in our simulations.  Further, as temperature is lowered, the average number of {\em random} modes needed to approximately reproduce the motion observed in inherent structure transitions remains constant, in stark contrast to the behavior seen when  $\langle N_{1/2}\rangle$ is calculated with the modes associated with the IS transitions.  In Fig.~\ref{fig:fall} we show that behavior illustrated in Fig.~\ref{fig:fhalftemp} does not depend on the choice of the threshold value $f$  used to define the degree of overlap between the modes and the true particle motion.  The behavior illustrated in Fig.~\ref{fig:fhalftemp} suggests that motion in mildly supercooled liquids proceeds along a successively smaller number of ``soft'' mode directions, a phenomena noted previously using a different definition of normal modes as hard spheres approach the jamming transition \cite{Brito-JSM2007}.


The behavior illustrated in Figs.~\ref{fig:disppart_count}-\ref{fig:displacements} is, upon first consideration, contradictory with the result presented in Fig.~\ref{fig:fhalftemp}.  In particular, Fig.~\ref{fig:disppart_count} shows that the motion that occurs during IS transitions becomes increasingly localized as temperature is decreased.  In Fig.~\ref{fig:displacements} we illustrate that, at least in one particular transition, it takes {\em more} modes to construct $\langle N_{1/2} \rangle$ for the case of a localized transition than for a delocalized one.  This fact can be quantified, and we show in Fig.~\ref{fig:fhalf_disppart} that this is true on average at any temperature. Thus it is surprising that  $\langle N_{1/2}\rangle$ decreases as temperature is lowered.  The resolution of this conflict lies in the fact that the curves of  $\langle N_{1/2}\rangle$ uniformly decrease for {\em all} types of transitions as temperature is lowered.  Clearly either something in the structure of the modes or the type of particle motion occurring in IS transitions (or both) changes as temperature is lowered such that the overlap between motion and modes increases in a uniform fashion independent of the degree of localization of the transition. Any changes in the mode structure are evidently subtle, at least in terms of the common metrics shown in Fig.~\ref{fig:dos}. In particular, since the temperature dependence of both the density of states and density of participation ratios of IS normal modes is very weak, differences must be found at the level of the small number of modes that actually contribute to  $\langle N_{1/2}\rangle$ at each temperature.

While the overall features of the various densities of states are largely insensitive to the lowering of the temperature, we can show that the properties of the modes that actually contribute to $\langle N_{1/2}\rangle$ do have properties that systematically change as temperature is varied. The contributing modes are essentially confined to the low frequency band in the neighborhood of the "boson" peak (with frequencies of $10 \tau^{-1}$ or smaller) at all temperatures studied.  Within this band, there is a systematic shift of the frequencies of contributing modes to lower values at lower temperatures, as illustrated in Fig.~\ref{fig:contrib_modes_freqs}.  Furthermore, for localized transitions the fraction of localized harmonic modes that contribute to $\langle N_{1/2}\rangle$ systematically increases as temperature is lowered, as shown in Fig.~\ref{fig:soft_contrib}. Thus, as temperature is decreased the modes that contribute to  $\langle N_{1/2}\rangle$ increasingly take on the character of the actual motion of particles during the IS transition.

\section{Conclusion}
\label{sec:conclusion}
The results of the previous section clearly show that dynamical heterogeneity in supercooled liquids, as exhibited by the motion of particles during IS trajectories, may be cleanly decomposed into a small number of harmonic normal modes associated with the inherent structures themselves. The fraction of Hessian eigenvectors needed to reproduce an average IS transition as quantified by $\langle N_{1/2}\rangle$ is on the order of two percent of the total number of modes and decreases as the liquid is increasingly supercooled.  This decrease in the number of soft mode directions that the system follows takes place even though IS transitions become more localized. Further, we have demonstrated that this property is a nontrivial feature of the actual pre- and post-IS mode structure and cannot be reproduced by modes obtained from independent inherent structures.

This study was clearly inspired by the work of Brito and Wyart, who found similar behavior in hard-spheres as the packing fraction is varied and the jamming transition is approached \cite{Brito-JSM2007, Brito-JCP2009}. Our work shows that this relationship between modes and dynamical heterogeneity is similar in thermal systems where inverse temperature plays the same role as density. Interestingly, we find that for an analogous range of relaxation times, even fewer modes are needed in our study to reproduce the quantity $\langle N_{1/2}\rangle$ than needed in the hard-sphere case. This is somewhat surprising because the statically defined IS normal modes are, in a sense, cruder than the dynamically defined modes of Brito and Wyart which are associated with the free energy (as opposed to potential energy) of the system.  It is still unclear if the results presented here can be taken as a feature of similarity between the jamming and glass transitions. In particular, given the importance of marginal stability in jamming, the depletion of mode directions as a basis for motion in jamming is natural. On the other hand the role of such mechanical consideration in thermal glass-forming systems is totally unclear. One interesting possibility hinted at by the lowest temperature data point in Fig.~\ref{fig:fhalftemp} is that the temperature dependence of  $\langle N_{1/2}\rangle$  plateaus at low temperature, perhaps close to the mode-coupling temperature of the system \cite{Krzakala-PRE2007,Mari-PRL2009,Parisi-RMP2010}.  While the quality of statistics we have is not good enough to draw such a conclusion, this possibility is worthy of future scrutiny. In particular, the jamming transition is accompanied by a {\em divergence} of the length scale associated with mode polarization and hence the directions associated with relaxation vanish at the transition \cite{Silbert-PRL2005}. In thermal systems, this behavior is absent or averted due to other processes, leading to a possible saturation of the number of soft directions that overlap with the true relaxation of the system.

Another avenue worthy of future study concerns the degree to which the results presented here are an indication of the primacy of soft modes in the relaxation of supercooled liquids.  In an interesting study, Ashton and Garrahan constructed models with mode structure by taking a standard lattice gas and connecting particles via springs with random force constants \cite{Ashton-EPJE2009}.  They showed that, within models where soft-modes are grafted onto kinetically constrained lattice gasses, the soft mode polarization will indeed congregate near vacancies but do not govern the motion of the defects, which instead move via local facilitated rules.  Thus, in this class of models, soft modes are {\em passive} markers of defect structure that do not themselves participate in relaxation. It would be most interesting to repeat the exercise of decomposing the inherent structures transitions in kinetically constrained models onto the normal modes of the Ashton-Garrahan models to test if the picture of modes as passive markers is compatible with the behavior exhibited in Fig.~\ref{fig:fhalftemp} and the previous work of Brito and Wyart. We reserve this investigation for a future study.

\begin{acknowledgments}
This research was performed on the following computing resources provided by the National Science Foundation (NSF): the PADS resource (NSF Grant No.~OCI-0821678) at at the Computation Institute, a joint institute of Argonne National Laboratory and the University of Chicago; the University of Chicago Computing Cooperative (UC3), supported in part by the Open Science Grid (NSF Grant No.~PHY-1148698); and the Extreme Science and Engineering Discovery Environment (XSEDE, NSF Grant No.~OCI-1053575). Simulations performed were organized by executing LAMMPS runs with the Swift parallel scripting language (NSF Grant No.~OCI-1148443) \cite{Wilde-Parallel2011}. G.M.H. and D.R.R. were supported by the NSF through Grant No. DGE-07-07425 and Grant No. CHE-0719089, respectively.
\end{acknowledgments}

\bibliography{modes_revised}

%Merlin.mbs v4.21 2009-07-09.
\begin{thebibliography}{10}%
\makeatletter
\providecommand \@ifxundefined [1]{%
 \ifx #1\undefined \expandafter \@firstoftwo
 \else \expandafter \@secondoftwo
\fi
}%
\providecommand \@ifnum [1]{%
 \ifnum #1\expandafter \@firstoftwo
 \else \expandafter \@secondoftwo
\fi
}%
\providecommand \enquote [1]{``#1''}%
\providecommand \bibnamefont  [1]{#1}%
\providecommand \bibfnamefont [1]{#1}%
\providecommand \citenamefont [1]{#1}%
\providecommand\href[0]{\@sanitize\@href}%
\providecommand\@href[1]{\endgroup\@@startlink{#1}\endgroup\@@href}%
\providecommand\@@href[1]{#1\@@endlink}%
\providecommand \@sanitize [0]{\begingroup\catcode`\&12\catcode`\#12\relax}%
\@ifxundefined \pdfoutput {\@firstoftwo}{%
 \@ifnum{\z@=\pdfoutput}{\@firstoftwo}{\@secondoftwo}%
}{%
 \providecommand\@@startlink[1]{\leavevmode\special{html:<a href="#1">}}%
 \providecommand\@@endlink[0]{\special{html:</a>}}%
}{%
 \providecommand\@@startlink[1]{%
  \leavevmode
  \pdfstartlink
   attr{/Border[0 0 1 ]/H/I/C[0 1 1]}%
   user{/Subtype/Link/A<</Type/Action/S/URI/URI(#1)>>}%
  \relax
 }%
 \providecommand\@@endlink[0]{\pdfendlink}%
}%
\providecommand \url  [0]{\begingroup\@sanitize \@url }%
\providecommand \@url [1]{\endgroup\@href {#1}{\urlprefix}}%
\providecommand \urlprefix [0]{URL }%
\providecommand \Eprint[0]{\href }%
\@ifxundefined \urlstyle {%
  \providecommand \doi [1]{doi:\discretionary{}{}{}#1}%
}{%
  \providecommand \doi [0]{doi:\discretionary{}{}{}\begingroup
  \urlstyle{rm}\Url }%
}%
\providecommand \doibase [0]{http://dx.doi.org/}%
\providecommand \Doi[1]{\href{\doibase#1}}%
\providecommand \selectlanguage [0]{\@gobble}%
\providecommand \bibinfo [0]{\@secondoftwo}%
\providecommand \bibfield [0]{\@secondoftwo}%
\providecommand \translation [1]{[#1]}%
\providecommand \BibitemOpen[0]{}%
\providecommand \bibitemStop [0]{}%
\providecommand \bibitemNoStop [0]{.\EOS\space}%
\providecommand \EOS [0]{\spacefactor3000\relax}%
\providecommand \BibitemShut [1]{\csname bibitem#1\endcsname}%
%</preamble>
\bibitem{Ediger-JPC1996}%
  \BibitemOpen
  \bibfield{author}{%
  \bibinfo {author} {\bibfnamefont{M.~D.}\ \bibnamefont{Ediger}}, \bibinfo
  {author} {\bibfnamefont{C.~A.}\ \bibnamefont{Angell}},\ and\ \bibinfo
  {author} {\bibfnamefont{S.~R.}\ \bibnamefont{Nagel}},\ }%
  \bibfield{journal}{%
  \bibinfo {journal} {J. Phys. Chem.}\ }%
  \textbf{\bibinfo {volume} {100}},\ \bibinfo {pages} {13200} (\bibinfo {year}
  {1996})\BibitemShut{NoStop}%
\bibitem{Debenedetti-Nature2001}%
  \BibitemOpen
  \bibfield{author}{%
  \bibinfo {author} {\bibfnamefont{P.}~\bibnamefont{Debenedetti}}, \bibinfo
  {author} {\bibfnamefont{F.}~\bibnamefont{Stillinger}}, \emph{et~al.},\ }%
  \bibfield{journal}{%
  \bibinfo {journal} {Nature}\ }%
  \textbf{\bibinfo {volume} {410}},\ \bibinfo {pages} {259} (\bibinfo {year}
  {2001})\BibitemShut{NoStop}%
\bibitem{Berthier-RMP2011}%
  \BibitemOpen
  \bibfield{author}{%
  \bibinfo {author} {\bibfnamefont{L.}~\bibnamefont{Berthier}}\ and\ \bibinfo
  {author} {\bibfnamefont{G.}~\bibnamefont{Biroli}},\ }%
  \bibfield{journal}{%
  \bibinfo {journal} {Rev. Mod. Phys.}\ }%
  \textbf{\bibinfo {volume} {83}},\ \bibinfo {pages} {587} (\bibinfo {year}
  {2011})\BibitemShut{NoStop}%
\bibitem{Ediger-ARPC2000}%
  \BibitemOpen
  \bibfield{author}{%
  \bibinfo {author} {\bibfnamefont{M.~D.}\ \bibnamefont{Ediger}},\ }%
  \bibfield{journal}{%
  \bibinfo {journal} {Annu. Rev. Phys. Chem.}\ }%
  \textbf{\bibinfo {volume} {51}},\ \bibinfo {pages} {99} (\bibinfo {year}
  {2000})\BibitemShut{NoStop}%
\bibitem{Berthier-Book2011}%
  \BibitemOpen
  \bibfield{author}{%
  \bibinfo {author} {\bibfnamefont{L.}~\bibnamefont{Berthier}}, \bibinfo
  {author} {\bibfnamefont{G.}~\bibnamefont{Biroli}}, \bibinfo {author}
  {\bibfnamefont{J.-P.}\ \bibnamefont{Bouchaud}}, \bibinfo {author}
  {\bibfnamefont{L.}~\bibnamefont{Cipelletti}},\ and\ \bibinfo {author}
  {\bibfnamefont{W.}~\bibnamefont{van Saarloos}},\ }%
  \emph{\bibinfo {title} {Dynamical heterogeneities in glasses, colloids, and
  granular media}}\ (\bibinfo {publisher} {Oxford University Press},\ \bibinfo
  {year} {2011})\BibitemShut{NoStop}%
\bibitem{Liu-Nature1998}%
  \BibitemOpen
  \bibfield{author}{%
  \bibinfo {author} {\bibfnamefont{A.~J.}\ \bibnamefont{Liu}}\ and\ \bibinfo
  {author} {\bibfnamefont{S.~R.}\ \bibnamefont{Nagel}},\ }%
  \bibfield{journal}{%
  \bibinfo {journal} {Nature}\ }%
  \textbf{\bibinfo {volume} {396}},\ \bibinfo {pages} {21} (\bibinfo {year}
  {1998})\BibitemShut{NoStop}%
\bibitem{Liu-ARCMP2010}%
  \BibitemOpen
  \bibfield{author}{%
  \bibinfo {author} {\bibfnamefont{A.~J.}\ \bibnamefont{Liu}}\ and\ \bibinfo
  {author} {\bibfnamefont{S.~R.}\ \bibnamefont{Nagel}},\ }%
  \bibfield{journal}{%
  \bibinfo {journal} {Ann. Rev. Cond. Mat. Phys.}\ }%
  \textbf{\bibinfo {volume} {1}},\ \bibinfo {pages} {347} (\bibinfo {year}
  {2010})\BibitemShut{NoStop}%
\bibitem{Zhang-Nature2009}%
  \BibitemOpen
  \bibfield{author}{%
  \bibinfo {author} {\bibfnamefont{Z.}~\bibnamefont{Zhang}}, \bibinfo {author}
  {\bibfnamefont{N.}~\bibnamefont{Xu}}, \bibinfo {author}
  {\bibfnamefont{D.~T.~N.}\ \bibnamefont{Chen}}, \bibinfo {author}
  {\bibfnamefont{P.}~\bibnamefont{Yunker}}, \bibinfo {author}
  {\bibfnamefont{A.~M.}\ \bibnamefont{Alsayed}}, \bibinfo {author}
  {\bibfnamefont{K.~B.}\ \bibnamefont{Aptowicz}}, \bibinfo {author}
  {\bibfnamefont{P.}~\bibnamefont{Habdas}}, \bibinfo {author}
  {\bibfnamefont{A.~J.}\ \bibnamefont{Liu}}, \bibinfo {author}
  {\bibfnamefont{S.~R.}\ \bibnamefont{Nagel}},\ and\ \bibinfo {author}
  {\bibfnamefont{A.~G.}\ \bibnamefont{Yodh}},\ }%
  \bibfield{journal}{%
  \bibinfo {journal} {Nature}\ }%
  \textbf{\bibinfo {volume} {459}},\ \bibinfo {pages} {230} (\bibinfo {year}
  {2009})\BibitemShut{NoStop}%
\bibitem{Haxton-PRE2011}%
  \BibitemOpen
  \bibfield{author}{%
  \bibinfo {author} {\bibfnamefont{T.~K.}\ \bibnamefont{Haxton}}, \bibinfo
  {author} {\bibfnamefont{M.}~\bibnamefont{Schmiedeberg}},\ and\ \bibinfo
  {author} {\bibfnamefont{A.~J.}\ \bibnamefont{Liu}},\ }%
  \bibfield{journal}{%
  \bibinfo {journal} {Phys. Rev. E.}\ }%
  \textbf{\bibinfo {volume} {83}},\ \bibinfo {pages} {031503} (\bibinfo {year}
  {2011})\BibitemShut{NoStop}%
\bibitem{Ikeda-PRL2012}%
  \BibitemOpen
  \bibfield{author}{%
  \bibinfo {author} {\bibfnamefont{A.}~\bibnamefont{Ikeda}}, \bibinfo {author}
  {\bibfnamefont{L.}~\bibnamefont{Berthier}},\ and\ \bibinfo {author}
  {\bibfnamefont{P.}~\bibnamefont{Sollich}},\ }%
  \bibfield{journal}{%
  \bibinfo {journal} {Phys. Rev. Lett.}\ }%
  \textbf{\bibinfo {volume} {109}},\ \bibinfo {pages} {018301} (\bibinfo {year}
  {2012})\BibitemShut{NoStop}%
\bibitem{Dauchot-PRL2005}%
  \BibitemOpen
  \bibfield{author}{%
  \bibinfo {author} {\bibfnamefont{O.}~\bibnamefont{Dauchot}}, \bibinfo
  {author} {\bibfnamefont{G.}~\bibnamefont{Marty}},\ and\ \bibinfo {author}
  {\bibfnamefont{G.}~\bibnamefont{Biroli}},\ }%
  \bibfield{journal}{%
  \bibinfo {journal} {Phys. Rev. Lett.}\ }%
  \textbf{\bibinfo {volume} {95}},\ \bibinfo {pages} {265701} (\bibinfo {year}
  {2005})\BibitemShut{NoStop}%
\bibitem{Keys-NatPhys2007}%
  \BibitemOpen
  \bibfield{author}{%
  \bibinfo {author} {\bibfnamefont{A.}~\bibnamefont{Keys}}, \bibinfo {author}
  {\bibfnamefont{A.}~\bibnamefont{Abate}}, \bibinfo {author}
  {\bibfnamefont{S.}~\bibnamefont{Glotzer}},\ and\ \bibinfo {author}
  {\bibfnamefont{D.}~\bibnamefont{Durian}},\ }%
  \bibfield{journal}{%
  \bibinfo {journal} {Nat. Phys.}\ }%
  \textbf{\bibinfo {volume} {3}},\ \bibinfo {pages} {260} (\bibinfo {year}
  {2007})\BibitemShut{NoStop}%
\bibitem{Dauchot-ChapterBook2011}%
  \BibitemOpen
  \bibfield{author}{%
  \bibinfo {author} {\bibfnamefont{O.}~\bibnamefont{Dauchot}}, \bibinfo
  {author} {\bibfnamefont{D.}~\bibnamefont{Durian}},\ and\ \bibinfo {author}
  {\bibfnamefont{M.}~\bibnamefont{van Hecke}},\ }%
  in\ \emph{\bibinfo {booktitle} {Dynamical Heterogeneities in Glasses,
  Colloids, and Granular Media}}\ (\bibinfo {publisher} {Oxford University
  Press},\ \bibinfo {year} {2011})\ pp.\ \bibinfo {pages}
  {110--151}\BibitemShut{NoStop}%
\bibitem{Silbert-PRL2005}%
  \BibitemOpen
  \bibfield{author}{%
  \bibinfo {author} {\bibfnamefont{L.}~\bibnamefont{Silbert}}, \bibinfo
  {author} {\bibfnamefont{A.}~\bibnamefont{Liu}},\ and\ \bibinfo {author}
  {\bibfnamefont{S.}~\bibnamefont{Nagel}},\ }%
  \bibfield{journal}{%
  \bibinfo {journal} {Phys. Rev. Lett.}\ }%
  \textbf{\bibinfo {volume} {95}},\ \bibinfo {pages} {098301} (\bibinfo {year}
  {2005})\BibitemShut{NoStop}%
\bibitem{Xu-PRL2007}%
  \BibitemOpen
  \bibfield{author}{%
  \bibinfo {author} {\bibfnamefont{N.}~\bibnamefont{Xu}}, \bibinfo {author}
  {\bibfnamefont{M.}~\bibnamefont{Wyart}}, \bibinfo {author}
  {\bibfnamefont{A.~J.}\ \bibnamefont{Liu}},\ and\ \bibinfo {author}
  {\bibfnamefont{S.~R.}\ \bibnamefont{Nagel}},\ }%
  \bibfield{journal}{%
  \bibinfo {journal} {Phys. Rev. Lett.}\ }%
  \textbf{\bibinfo {volume} {98}},\ \bibinfo {pages} {175502} (\bibinfo {year}
  {2007})\BibitemShut{NoStop}%
\bibitem{Xu-PRL2009}%
  \BibitemOpen
  \bibfield{author}{%
  \bibinfo {author} {\bibfnamefont{N.}~\bibnamefont{Xu}}, \bibinfo {author}
  {\bibfnamefont{V.}~\bibnamefont{Vitelli}}, \bibinfo {author}
  {\bibfnamefont{M.}~\bibnamefont{Wyart}}, \bibinfo {author}
  {\bibfnamefont{A.~J.}\ \bibnamefont{Liu}},\ and\ \bibinfo {author}
  {\bibfnamefont{S.~R.}\ \bibnamefont{Nagel}},\ }%
  \bibfield{journal}{%
  \bibinfo {journal} {Phys. Rev. Lett.}\ }%
  \textbf{\bibinfo {volume} {102}},\ \bibinfo {pages} {038001} (\bibinfo {year}
  {2009})\BibitemShut{NoStop}%
\bibitem{Goodrich-PRL2012}%
  \BibitemOpen
  \bibfield{author}{%
  \bibinfo {author} {\bibfnamefont{C.~P.}\ \bibnamefont{Goodrich}}, \bibinfo
  {author} {\bibfnamefont{A.~J.}\ \bibnamefont{Liu}},\ and\ \bibinfo {author}
  {\bibfnamefont{S.~R.}\ \bibnamefont{Nagel}},\ }%
  \bibfield{journal}{%
  \bibinfo {journal} {Phys. Rev. Lett.}\ }%
  \textbf{\bibinfo {volume} {109}},\ \bibinfo {pages} {095704} (\bibinfo {year}
  {2012})\BibitemShut{NoStop}%
\bibitem{Schober-PRB1991}%
  \BibitemOpen
  \bibfield{author}{%
  \bibinfo {author} {\bibfnamefont{H.~R.}\ \bibnamefont{Schober}}\ and\
  \bibinfo {author} {\bibfnamefont{B.~B.}\ \bibnamefont{Laird}},\ }%
  \bibfield{journal}{%
  \bibinfo {journal} {Phys. Rev. B}\ }%
  \textbf{\bibinfo {volume} {44}},\ \bibinfo {pages} {6746} (\bibinfo {year}
  {1991})\BibitemShut{NoStop}%
\bibitem{Laird-PRL1991}%
  \BibitemOpen
  \bibfield{author}{%
  \bibinfo {author} {\bibfnamefont{B.~B.}\ \bibnamefont{Laird}}\ and\ \bibinfo
  {author} {\bibfnamefont{H.~R.}\ \bibnamefont{Schober}},\ }%
  \bibfield{journal}{%
  \bibinfo {journal} {Phys. Rev. Lett.}\ }%
  \textbf{\bibinfo {volume} {66}},\ \bibinfo {pages} {636} (\bibinfo {year}
  {1991})\BibitemShut{NoStop}%
\bibitem{Biroli-PRL2006}%
  \BibitemOpen
  \bibfield{author}{%
  \bibinfo {author} {\bibfnamefont{G.}~\bibnamefont{Biroli}}, \bibinfo {author}
  {\bibfnamefont{J.-P.}\ \bibnamefont{Bouchaud}}, \bibinfo {author}
  {\bibfnamefont{K.}~\bibnamefont{Miyazaki}},\ and\ \bibinfo {author}
  {\bibfnamefont{D.~R.}\ \bibnamefont{Reichman}},\ }%
  \bibfield{journal}{%
  \bibinfo {journal} {Phys. Rev. Lett.}\ }%
  \textbf{\bibinfo {volume} {97}},\ \bibinfo {pages} {195701} (\bibinfo {year}
  {2006})\BibitemShut{NoStop}%
\bibitem{Krzakala-PRE2007}%
  \BibitemOpen
  \bibfield{author}{%
  \bibinfo {author} {\bibfnamefont{F.}~\bibnamefont{Krzakala}}\ and\ \bibinfo
  {author} {\bibfnamefont{J.}~\bibnamefont{Kurchan}},\ }%
  \bibfield{journal}{%
  \bibinfo {journal} {Phys. Rev. E}\ }%
  \textbf{\bibinfo {volume} {76}},\ \bibinfo {pages} {021122} (\bibinfo {year}
  {2007})\BibitemShut{NoStop}%
\bibitem{Kurchan-JPA1996}%
  \BibitemOpen
  \bibfield{author}{%
  \bibinfo {author} {\bibfnamefont{J.}~\bibnamefont{Kurchan}}\ and\ \bibinfo
  {author} {\bibfnamefont{L.}~\bibnamefont{Laloux}},\ }%
  \bibfield{journal}{%
  \bibinfo {journal} {J. Phys. A: Math. Gen}\ }%
  \textbf{\bibinfo {volume} {29}},\ \bibinfo {pages} {1929} (\bibinfo {year}
  {1996})\BibitemShut{NoStop}%
\bibitem{Brito-EPL2006}%
  \BibitemOpen
  \bibfield{author}{%
  \bibinfo {author} {\bibfnamefont{C.}~\bibnamefont{Brito}}\ and\ \bibinfo
  {author} {\bibfnamefont{M.}~\bibnamefont{Wyart}},\ }%
  \bibfield{journal}{%
  \bibinfo {journal} {EPL}\ }%
  \textbf{\bibinfo {volume} {76}},\ \bibinfo {pages} {149} (\bibinfo {year}
  {2006})\BibitemShut{NoStop}%
\bibitem{Brito-JCP2009}%
  \BibitemOpen
  \bibfield{author}{%
  \bibinfo {author} {\bibfnamefont{C.}~\bibnamefont{Brito}}\ and\ \bibinfo
  {author} {\bibfnamefont{M.}~\bibnamefont{Wyart}},\ }%
  \bibfield{journal}{%
  \bibinfo {journal} {J. Chem. Phys.}\ }%
  \textbf{\bibinfo {volume} {131}},\ \bibinfo {pages} {024504} (\bibinfo {year}
  {2009})\BibitemShut{NoStop}%
\bibitem{WidmerCooper-PRL2004}%
  \BibitemOpen
  \bibfield{author}{%
  \bibinfo {author} {\bibfnamefont{A.}~\bibnamefont{Widmer-Cooper}}, \bibinfo
  {author} {\bibfnamefont{P.}~\bibnamefont{Harrowell}},\ and\ \bibinfo {author}
  {\bibfnamefont{H.}~\bibnamefont{Fynewever}},\ }%
  \bibfield{journal}{%
  \bibinfo {journal} {Phys. Rev. Lett.}\ }%
  \textbf{\bibinfo {volume} {93}},\ \bibinfo {pages} {135701} (\bibinfo {year}
  {2004})\BibitemShut{NoStop}%
\bibitem{WidmerCooper-JPCM2005}%
  \BibitemOpen
  \bibfield{author}{%
  \bibinfo {author} {\bibfnamefont{A.}~\bibnamefont{Widmer-Cooper}}\ and\
  \bibinfo {author} {\bibfnamefont{P.}~\bibnamefont{Harrowell}},\ }%
  \bibfield{journal}{%
  \bibinfo {journal} {J. Phys: Cond. Mat.}\ }%
  \textbf{\bibinfo {volume} {17}},\ \bibinfo {pages} {S4025} (\bibinfo {year}
  {2005})\BibitemShut{NoStop}%
\bibitem{WidmerCooper-PRL2006}%
  \BibitemOpen
  \bibfield{author}{%
  \bibinfo {author} {\bibfnamefont{A.}~\bibnamefont{Widmer-Cooper}}\ and\
  \bibinfo {author} {\bibfnamefont{P.}~\bibnamefont{Harrowell}},\ }%
  \bibfield{journal}{%
  \bibinfo {journal} {Phys. Rev. Lett.}\ }%
  \textbf{\bibinfo {volume} {96}},\ \bibinfo {pages} {185701} (\bibinfo {year}
  {2006})\BibitemShut{NoStop}%
\bibitem{WidmerCooper-NatPhys2008}%
  \BibitemOpen
  \bibfield{author}{%
  \bibinfo {author} {\bibfnamefont{A.}~\bibnamefont{Widmer-Cooper}}, \bibinfo
  {author} {\bibfnamefont{H.}~\bibnamefont{Perry}}, \bibinfo {author}
  {\bibfnamefont{P.}~\bibnamefont{Harrowell}},\ and\ \bibinfo {author}
  {\bibfnamefont{D.~R.}\ \bibnamefont{Reichman}},\ }%
  \bibfield{journal}{%
  \bibinfo {journal} {Nat. Phys.}\ }%
  \textbf{\bibinfo {volume} {4}},\ \bibinfo {pages} {711} (\bibinfo {year}
  {2008})\BibitemShut{NoStop}%
\bibitem{WidmerCooper-JCP2009}%
  \BibitemOpen
  \bibfield{author}{%
  \bibinfo {author} {\bibfnamefont{A.}~\bibnamefont{Widmer-Cooper}}, \bibinfo
  {author} {\bibfnamefont{H.}~\bibnamefont{Perry}}, \bibinfo {author}
  {\bibfnamefont{P.}~\bibnamefont{Harrowell}},\ and\ \bibinfo {author}
  {\bibfnamefont{D.~R.}\ \bibnamefont{Reichman}},\ }%
  \bibfield{journal}{%
  \bibinfo {journal} {J. Chem. Phys.}\ }%
  \textbf{\bibinfo {volume} {131}},\ \bibinfo {pages} {194508} (\bibinfo {year}
  {2009})\BibitemShut{NoStop}%
\bibitem{Candelier-PRL2010}%
  \BibitemOpen
  \bibfield{author}{%
  \bibinfo {author} {\bibfnamefont{R.}~\bibnamefont{Candelier}}, \bibinfo
  {author} {\bibfnamefont{A.}~\bibnamefont{Widmer-Cooper}}, \bibinfo {author}
  {\bibfnamefont{J.~K.}\ \bibnamefont{Kummerfeld}}, \bibinfo {author}
  {\bibfnamefont{O.}~\bibnamefont{Dauchot}}, \bibinfo {author}
  {\bibfnamefont{G.}~\bibnamefont{Biroli}}, \bibinfo {author}
  {\bibfnamefont{P.}~\bibnamefont{Harrowell}},\ and\ \bibinfo {author}
  {\bibfnamefont{D.~R.}\ \bibnamefont{Reichman}},\ }%
  \bibfield{journal}{%
  \bibinfo {journal} {Phys. Rev. Lett.}\ }%
  \textbf{\bibinfo {volume} {105}},\ \bibinfo {pages} {135702} (\bibinfo {year}
  {2010})\BibitemShut{NoStop}%
\bibitem{Xu-EPL2010}%
  \BibitemOpen
  \bibfield{author}{%
  \bibinfo {author} {\bibfnamefont{N.}~\bibnamefont{Xu}}, \bibinfo {author}
  {\bibfnamefont{V.}~\bibnamefont{Vitelli}}, \bibinfo {author}
  {\bibfnamefont{A.~J.}\ \bibnamefont{Liu}},\ and\ \bibinfo {author}
  {\bibfnamefont{S.~R.}\ \bibnamefont{Nagel}},\ }%
  \bibfield{journal}{%
  \bibinfo {journal} {Euro. Phys. Lett.}\ }%
  \textbf{\bibinfo {volume} {90}},\ \bibinfo {pages} {56001} (\bibinfo {year}
  {2010})\BibitemShut{NoStop}%
\bibitem{Brito-SoftMat2010}%
  \BibitemOpen
  \bibfield{author}{%
  \bibinfo {author} {\bibfnamefont{C.}~\bibnamefont{Brito}}, \bibinfo {author}
  {\bibfnamefont{O.}~\bibnamefont{Dauchot}}, \bibinfo {author}
  {\bibfnamefont{G.}~\bibnamefont{Biroli}},\ and\ \bibinfo {author}
  {\bibfnamefont{J.-P.}\ \bibnamefont{Bouchaud}},\ }%
  \bibfield{journal}{%
  \bibinfo {journal} {Soft Matter}\ }%
  \textbf{\bibinfo {volume} {6}},\ \bibinfo {pages} {3013} (\bibinfo {year}
  {2010})\BibitemShut{NoStop}%
\bibitem{Ghosh-PRL2010}%
  \BibitemOpen
  \bibfield{author}{%
  \bibinfo {author} {\bibfnamefont{A.}~\bibnamefont{Ghosh}}, \bibinfo {author}
  {\bibfnamefont{V.}~\bibnamefont{Chikkadi}}, \bibinfo {author}
  {\bibfnamefont{P.}~\bibnamefont{Schall}}, \bibinfo {author}
  {\bibfnamefont{J.}~\bibnamefont{Kurchan}},\ and\ \bibinfo {author}
  {\bibfnamefont{D.}~\bibnamefont{Bonn}},\ }%
  \bibfield{journal}{%
  \bibinfo {journal} {Phys. Rev. Lett.}\ }%
  \textbf{\bibinfo {volume} {104}},\ \bibinfo {pages} {248305} (\bibinfo {year}
  {2010})\BibitemShut{NoStop}%
\bibitem{Chen-PRL2010}%
  \BibitemOpen
  \bibfield{author}{%
  \bibinfo {author} {\bibfnamefont{K.}~\bibnamefont{Chen}}, \bibinfo {author}
  {\bibfnamefont{W.~G.}\ \bibnamefont{Ellenbroek}}, \bibinfo {author}
  {\bibfnamefont{Z.}~\bibnamefont{Zhang}}, \bibinfo {author}
  {\bibfnamefont{D.~T.~N.}\ \bibnamefont{Chen}}, \bibinfo {author}
  {\bibfnamefont{P.~J.}\ \bibnamefont{Yunker}}, \bibinfo {author}
  {\bibfnamefont{S.}~\bibnamefont{Henkes}}, \bibinfo {author}
  {\bibfnamefont{C.}~\bibnamefont{Brito}}, \bibinfo {author}
  {\bibfnamefont{O.}~\bibnamefont{Dauchot}}, \bibinfo {author}
  {\bibfnamefont{W.}~\bibnamefont{van Saarloos}}, \bibinfo {author}
  {\bibfnamefont{A.~J.}\ \bibnamefont{Liu}},\ and\ \bibinfo {author}
  {\bibfnamefont{A.~G.}\ \bibnamefont{Yodh}},\ }%
  \bibfield{journal}{%
  \bibinfo {journal} {Phys. Rev. Lett.}\ }%
  \textbf{\bibinfo {volume} {105}},\ \bibinfo {pages} {025501} (\bibinfo {year}
  {2010})\BibitemShut{NoStop}%
\bibitem{Chen-PRL2011}%
  \BibitemOpen
  \bibfield{author}{%
  \bibinfo {author} {\bibfnamefont{K.}~\bibnamefont{Chen}}, \bibinfo {author}
  {\bibfnamefont{M.~L.}\ \bibnamefont{Manning}}, \bibinfo {author}
  {\bibfnamefont{P.~J.}\ \bibnamefont{Yunker}}, \bibinfo {author}
  {\bibfnamefont{W.~G.}\ \bibnamefont{Ellenbroek}}, \bibinfo {author}
  {\bibfnamefont{Z.}~\bibnamefont{Zhang}}, \bibinfo {author}
  {\bibfnamefont{A.~J.}\ \bibnamefont{Liu}},\ and\ \bibinfo {author}
  {\bibfnamefont{A.~G.}\ \bibnamefont{Yodh}},\ }%
  \bibfield{journal}{%
  \bibinfo {journal} {Phys. Rev. Lett.}\ }%
  \textbf{\bibinfo {volume} {107}},\ \bibinfo {pages} {108301} (\bibinfo {year}
  {2011})\BibitemShut{NoStop}%
\bibitem{Brito-JSM2007}%
  \BibitemOpen
  \bibfield{author}{%
  \bibinfo {author} {\bibfnamefont{C.}~\bibnamefont{Brito}}\ and\ \bibinfo
  {author} {\bibfnamefont{M.}~\bibnamefont{Wyart}},\ }%
  \bibfield{journal}{%
  \bibinfo {journal} {J. Stat. Mech. Theor. Exp.}\ }%
  \textbf{\bibinfo {volume} {2007}},\ \bibinfo {pages} {L08003} (\bibinfo
  {year} {2007})\BibitemShut{NoStop}%
\bibitem{Bruning-JPCM2009}%
  \BibitemOpen
  \bibfield{author}{%
  \bibinfo {author} {\bibfnamefont{R.}~\bibnamefont{Br{\"u}ning}}, \bibinfo
  {author} {\bibfnamefont{D.~A.}\ \bibnamefont{St-Onge}}, \bibinfo {author}
  {\bibfnamefont{S.}~\bibnamefont{Patterson}},\ and\ \bibinfo {author}
  {\bibfnamefont{W.}~\bibnamefont{Kob}},\ }%
  \bibfield{journal}{%
  \bibinfo {journal} {J. Phys. Cond. Mat.}\ }%
  \textbf{\bibinfo {volume} {21}},\ \bibinfo {pages} {035117} (\bibinfo {year}
  {2009})\BibitemShut{NoStop}%
\bibitem{Karmakar-PRE2012}%
  \BibitemOpen
  \bibfield{author}{%
  \bibinfo {author} {\bibfnamefont{S.}~\bibnamefont{{Karmakar}}}\ and\ \bibinfo
  {author} {\bibfnamefont{I.}~\bibnamefont{{Procaccia}}},\ }%
  \bibfield{journal}{%
  \bibinfo {journal} {Phys. Rev. E}\ }%
  \textbf{\bibinfo {volume} {86}},\ \bibinfo {pages} {061502} (\bibinfo {year}
  {2012})\BibitemShut{NoStop}%
\bibitem{Kob-PRL1994}%
  \BibitemOpen
  \bibfield{author}{%
  \bibinfo {author} {\bibfnamefont{W.}~\bibnamefont{Kob}}\ and\ \bibinfo
  {author} {\bibfnamefont{H.}~\bibnamefont{Andersen}},\ }%
  \bibfield{journal}{%
  \bibinfo {journal} {Phys. Rev. Lett.}\ }%
  \textbf{\bibinfo {volume} {73}},\ \bibinfo {pages} {1376} (\bibinfo {year}
  {1994})\BibitemShut{NoStop}%
\bibitem{Stillinger-PRA1982}%
  \BibitemOpen
  \bibfield{author}{%
  \bibinfo {author} {\bibfnamefont{F.}~\bibnamefont{Stillinger}}\ and\ \bibinfo
  {author} {\bibfnamefont{T.}~\bibnamefont{Weber}},\ }%
  \bibfield{journal}{%
  \bibinfo {journal} {Phys. Rev. A}\ }%
  \textbf{\bibinfo {volume} {25}},\ \bibinfo {pages} {978} (\bibinfo {year}
  {1982})\BibitemShut{NoStop}%
\bibitem{Weber-JCP1984}%
  \BibitemOpen
  \bibfield{author}{%
  \bibinfo {author} {\bibfnamefont{T.}~\bibnamefont{Weber}}\ and\ \bibinfo
  {author} {\bibfnamefont{F.}~\bibnamefont{Stillinger}},\ }%
  \bibfield{journal}{%
  \bibinfo {journal} {J. Chem. Phys.}\ }%
  \textbf{\bibinfo {volume} {81}},\ \bibinfo {pages} {5089} (\bibinfo {year}
  {1984})\BibitemShut{NoStop}%
\bibitem{Weber-PRB1985}%
  \BibitemOpen
  \bibfield{author}{%
  \bibinfo {author} {\bibfnamefont{T.~A.}\ \bibnamefont{Weber}}\ and\ \bibinfo
  {author} {\bibfnamefont{F.~H.}\ \bibnamefont{Stillinger}},\ }%
  \bibfield{journal}{%
  \bibinfo {journal} {Phys. Rev. B}\ }%
  \textbf{\bibinfo {volume} {31}},\ \bibinfo {pages} {1954} (\bibinfo {year}
  {1985})\BibitemShut{NoStop}%
\bibitem{Sastry-Nature1998}%
  \BibitemOpen
  \bibfield{author}{%
  \bibinfo {author} {\bibfnamefont{S.}~\bibnamefont{Sastry}}, \bibinfo {author}
  {\bibfnamefont{P.}~\bibnamefont{Debenedetti}},\ and\ \bibinfo {author}
  {\bibfnamefont{F.}~\bibnamefont{Stillinger}},\ }%
  \bibfield{journal}{%
  \bibinfo {journal} {Nature}\ }%
  \textbf{\bibinfo {volume} {393}},\ \bibinfo {pages} {554} (\bibinfo {year}
  {1998})\BibitemShut{NoStop}%
\bibitem{Schroder-JCP2000}%
  \BibitemOpen
  \bibfield{author}{%
  \bibinfo {author} {\bibfnamefont{T.~B.}\ \bibnamefont{Schr{\o}der}}, \bibinfo
  {author} {\bibfnamefont{S.}~\bibnamefont{Sastry}}, \bibinfo {author}
  {\bibfnamefont{J.~C.}\ \bibnamefont{Dyre}},\ and\ \bibinfo {author}
  {\bibfnamefont{S.~C.}\ \bibnamefont{Glotzer}},\ }%
  \bibfield{journal}{%
  \bibinfo {journal} {J. Chem. Phys.}\ }%
  \textbf{\bibinfo {volume} {112}},\ \bibinfo {pages} {9834} (\bibinfo {year}
  {2000})\BibitemShut{NoStop}%
\bibitem{Keys-PRX2011}%
  \BibitemOpen
  \bibfield{author}{%
  \bibinfo {author} {\bibfnamefont{A.~S.}\ \bibnamefont{Keys}}, \bibinfo
  {author} {\bibfnamefont{L.~O.}\ \bibnamefont{Hedges}}, \bibinfo {author}
  {\bibfnamefont{J.~P.}\ \bibnamefont{Garrahan}}, \bibinfo {author}
  {\bibfnamefont{S.~C.}\ \bibnamefont{Glotzer}},\ and\ \bibinfo {author}
  {\bibfnamefont{D.}~\bibnamefont{Chandler}},\ }%
  \bibfield{journal}{%
  \bibinfo {journal} {Phys. Rev. X}\ }%
  \textbf{\bibinfo {volume} {1}},\ \bibinfo {pages} {021013} (\bibinfo {year}
  {2011})\BibitemShut{NoStop}%
\bibitem{Plimpton-JCP1995}%
  \BibitemOpen
  \bibfield{author}{%
  \bibinfo {author} {\bibfnamefont{S.}~\bibnamefont{Plimpton}},\ }%
  \bibfield{journal}{%
  \bibinfo {journal} {J. Comp. Phys.}\ }%
  \textbf{\bibinfo {volume} {117}},\ \bibinfo {pages} {1} (\bibinfo {year}
  {1995}),\ \url{http://lammps.sandia.gov}\BibitemShut{NoStop}%
\bibitem{Doliwa-PRE2003a}%
  \BibitemOpen
  \bibfield{author}{%
  \bibinfo {author} {\bibfnamefont{B.}~\bibnamefont{Doliwa}}\ and\ \bibinfo
  {author} {\bibfnamefont{A.}~\bibnamefont{Heuer}},\ }%
  \bibfield{journal}{%
  \bibinfo {journal} {Phys. Rev. E}\ }%
  \textbf{\bibinfo {volume} {67}},\ \bibinfo {pages} {030501} (\bibinfo {year}
  {2003})\BibitemShut{NoStop}%
\bibitem{Doliwa-PRE2003b}%
  \BibitemOpen
  \bibfield{author}{%
  \bibinfo {author} {\bibfnamefont{B.}~\bibnamefont{Doliwa}}\ and\ \bibinfo
  {author} {\bibfnamefont{A.}~\bibnamefont{Heuer}},\ }%
  \bibfield{journal}{%
  \bibinfo {journal} {Phys. Rev. E}\ }%
  \textbf{\bibinfo {volume} {67}},\ \bibinfo {pages} {031506} (\bibinfo {year}
  {2003})\BibitemShut{NoStop}%
\bibitem{Denny-PRL2003}%
  \BibitemOpen
  \bibfield{author}{%
  \bibinfo {author} {\bibfnamefont{R.~A.}\ \bibnamefont{Denny}}, \bibinfo
  {author} {\bibfnamefont{D.~R.}\ \bibnamefont{Reichman}},\ and\ \bibinfo
  {author} {\bibfnamefont{J.-P.}\ \bibnamefont{Bouchaud}},\ }%
  \bibfield{journal}{%
  \bibinfo {journal} {Phys. Rev. Lett.}\ }%
  \textbf{\bibinfo {volume} {90}},\ \bibinfo {pages} {025503} (\bibinfo {year}
  {2003})\BibitemShut{NoStop}%
\bibitem{Note1}%
  \BibitemOpen
  \bibinfo {note} {In fact, all quantities computed which compare modes with
  inherent structure transition displacement vectors were identical within
  error bars if computed using pre- and post-transition modes, and hence when
  calculating the values in Figs.~\ref {fig:fall}-\ref {fig:soft_contrib} we
  average both sets of data for increased statistical
  accuracy.}\BibitemShut{Stop}%
\bibitem{Efron-1993}%
  \BibitemOpen
  \bibfield{author}{%
  \bibinfo {author} {\bibfnamefont{B.}~\bibnamefont{Efron}}\ and\ \bibinfo
  {author} {\bibfnamefont{R.}~\bibnamefont{Tibshirani}},\ }%
  \emph{\bibinfo {title} {An introduction to the bootstrap}},\ Vol.~\bibinfo
  {volume} {57}\ (\bibinfo {publisher} {Chapman \& Hall/CRC},\ \bibinfo {year}
  {1993})\BibitemShut{NoStop}%
\bibitem{Mari-PRL2009}%
  \BibitemOpen
  \bibfield{author}{%
  \bibinfo {author} {\bibfnamefont{R.}~\bibnamefont{Mari}}, \bibinfo {author}
  {\bibfnamefont{F.}~\bibnamefont{Krzakala}},\ and\ \bibinfo {author}
  {\bibfnamefont{J.}~\bibnamefont{Kurchan}},\ }%
  \bibfield{journal}{%
  \bibinfo {journal} {Phys. Rev. Lett.}\ }%
  \textbf{\bibinfo {volume} {103}},\ \bibinfo {pages} {025701} (\bibinfo {year}
  {2009})\BibitemShut{NoStop}%
\bibitem{Parisi-RMP2010}%
  \BibitemOpen
  \bibfield{author}{%
  \bibinfo {author} {\bibfnamefont{G.}~\bibnamefont{Parisi}}\ and\ \bibinfo
  {author} {\bibfnamefont{F.}~\bibnamefont{Zamponi}},\ }%
  \bibfield{journal}{%
  \bibinfo {journal} {Rev. Mod. Phys.}\ }%
  \textbf{\bibinfo {volume} {82}},\ \bibinfo {pages} {789} (\bibinfo {year}
  {2010})\BibitemShut{NoStop}%
\bibitem{Ashton-EPJE2009}%
  \BibitemOpen
  \bibfield{author}{%
  \bibinfo {author} {\bibfnamefont{D.~J.}\ \bibnamefont{Ashton}}\ and\ \bibinfo
  {author} {\bibfnamefont{J.~P.}\ \bibnamefont{Garrahan}},\ }%
  \bibfield{journal}{%
  \bibinfo {journal} {Euro. Phys. J. E}\ }%
  \textbf{\bibinfo {volume} {30}},\ \bibinfo {pages} {303} (\bibinfo {year}
  {2009})\BibitemShut{NoStop}%
\bibitem{Wilde-Parallel2011}%
  \BibitemOpen
  \bibfield{author}{%
  \bibinfo {author} {\bibfnamefont{M.}~\bibnamefont{Wilde}}, \bibinfo {author}
  {\bibfnamefont{M.}~\bibnamefont{Hategan}}, \bibinfo {author}
  {\bibfnamefont{J.~M.}\ \bibnamefont{Wozniak}}, \bibinfo {author}
  {\bibfnamefont{B.}~\bibnamefont{Clifford}}, \bibinfo {author}
  {\bibfnamefont{D.~S.}\ \bibnamefont{Katz}},\ and\ \bibinfo {author}
  {\bibfnamefont{I.}~\bibnamefont{Foster}},\ }%
  \bibfield{journal}{%
  \bibinfo {journal} {Parallel Comput.}\ }%
  \textbf{\bibinfo {volume} {37}},\ \bibinfo {pages} {633 } (\bibinfo {year}
  {2011})\BibitemShut{NoStop}%
\end{thebibliography}%

\end{document}